\documentclass{article}

\usepackage[preprint]{neurips_2025}

\usepackage[utf8]{inputenc} 
\usepackage[T1]{fontenc}    
\usepackage{hyperref}       
\usepackage{url}            
\usepackage{booktabs}       
\usepackage{amsfonts}       
\usepackage{nicefrac}       
\usepackage{microtype}      
\usepackage{xcolor}         
\usepackage{graphicx}
\usepackage{array} 
\usepackage{enumitem}
\usepackage{listings}
\usepackage{tcolorbox}
\tcbuselibrary{breakable, skins}
\usepackage{fontawesome}
\usepackage{float}

\newtcolorbox[auto counter, number within=section]{methodbox}[2][]{%
  colback=white, 
  colframe=teal!80!green!80!black,  
  width=\textwidth,
  arc=2mm, 
  boxrule=0.5mm, 
  title={\normalsize\faWrench\hspace{0.5em}#2}, 
  breakable,
  fonttitle=\bfseries\Large, 
  fontupper=\small,
  #1
}

\newtcolorbox[auto counter, number within=section]{promptbox}[2][]{%
  colback=white, 
  colframe=purple!70!blue!80!black,  
  width=\textwidth,
  arc=2mm, 
  boxrule=0.5mm, 
  title={\normalsize\faInfoCircle\hspace{0.5em}#2},
  breakable,
  fonttitle=\bfseries\Large, 
  fontupper=\small, 
  drop shadow southeast, 
  #1
}

\newtcolorbox[auto counter, number within=section]{purposebox}[2][]{%
  colback=white, 
  colframe=blue!50!black, 
  width=\textwidth,
  arc=2mm, 
  boxrule=0.5mm, 
  title={\normalsize\faCompass\hspace{0.5em}#2},
  breakable, 
  fonttitle=\bfseries\Large, 
  fontupper=\small, 
  drop shadow southeast, 
  #1
}

\definecolor{cvprblue}{rgb}{0.21,0.49,0.74}
\definecolor{maroon}{RGB}{255, 20, 147}
\hypersetup{
    colorlinks=true,            
    linkcolor=maroon,             
    filecolor=magenta,         
    urlcolor=cyan,              
    citecolor=cvprblue,             
    pdftitle={Overleaf Example},
    pdfpagemode=FullScreen,
}

\title{Anemoi: A Semi-Centralized Multi-agent System Based on Agent-to-Agent Communication MCP server from Coral Protocol}

\author{%
  \textbf{Xinxing Ren}$^{1,2,*}$ \quad
  \textbf{Caelum Forder}$^{1,*}$ \quad
  \textbf{Qianbo Zang}$^{3,*}$ \quad
  \textbf{Ahsen Tahir}$^{1,5}$ \\
  \textbf{Roman J. Georgio}$^{1}$ \quad
  \textbf{Suman Deb}$^{1}$ \quad
  \textbf{Peter Carroll}$^{1}$ \quad
  \textbf{{\"O}nder G{\"u}rcan}$^{1,\dagger}$ \quad
  \textbf{Zekun Guo}$^{4,\dagger}$ \\[6pt]
  $^{1}$ Coral Protocol \quad
  $^{2}$ Brunel University of London \quad
  $^{3}$ SnT, Université du Luxembourg \\
  $^{4}$ University of Hull \quad
  $^{5}$ National University of Computer and Emerging Sciences \\[6pt]
  $^{*}$ Equal contribution \quad
  $^{\dagger}$ Co-corresponding authors
}

\begin{document}

\maketitle

\begin{abstract}
Recent advances in generalist multi-agent systems (MAS) have largely followed a context-engineering plus centralized paradigm, where a planner agent coordinates multiple worker agents through unidirectional prompt passing. While effective under strong planner models, this design suffers from two critical limitations: (1) strong dependency on the planner’s capability, which leads to degraded performance when a smaller LLM powers the planner; and (2) limited inter-agent communication, where collaboration relies on prompt concatenation rather than genuine refinement through structured discussions. To address these challenges, we propose \textbf{Anemoi}, a semi-centralized MAS built on the Agent-to-Agent (A2A) communication MCP server from \textbf{Coral Protocol}. Unlike traditional designs, Anemoi enables structured and direct inter-agent collaboration, allowing all agents to monitor progress, assess results, identify bottlenecks, and propose refinements in real time. This paradigm reduces reliance on a single planner, supports adaptive plan updates, and minimizes redundant context passing, resulting in more scalable execution. Evaluated on the GAIA benchmark, Anemoi achieved \textbf{52.73\%} accuracy with a small LLM (\texttt{GPT-4.1-mini}) as the planner, surpassing the strongest open-source baseline OWL (43.63\%) by \textbf{+9.09\%} under identical LLM settings. Our implementation is publicly available at \url{https://github.com/Coral-Protocol/Anemoi}. 
\end{abstract}

\section{Introduction}

\begin{figure}[h]
    \centering
    \includegraphics[width=1.0\linewidth]{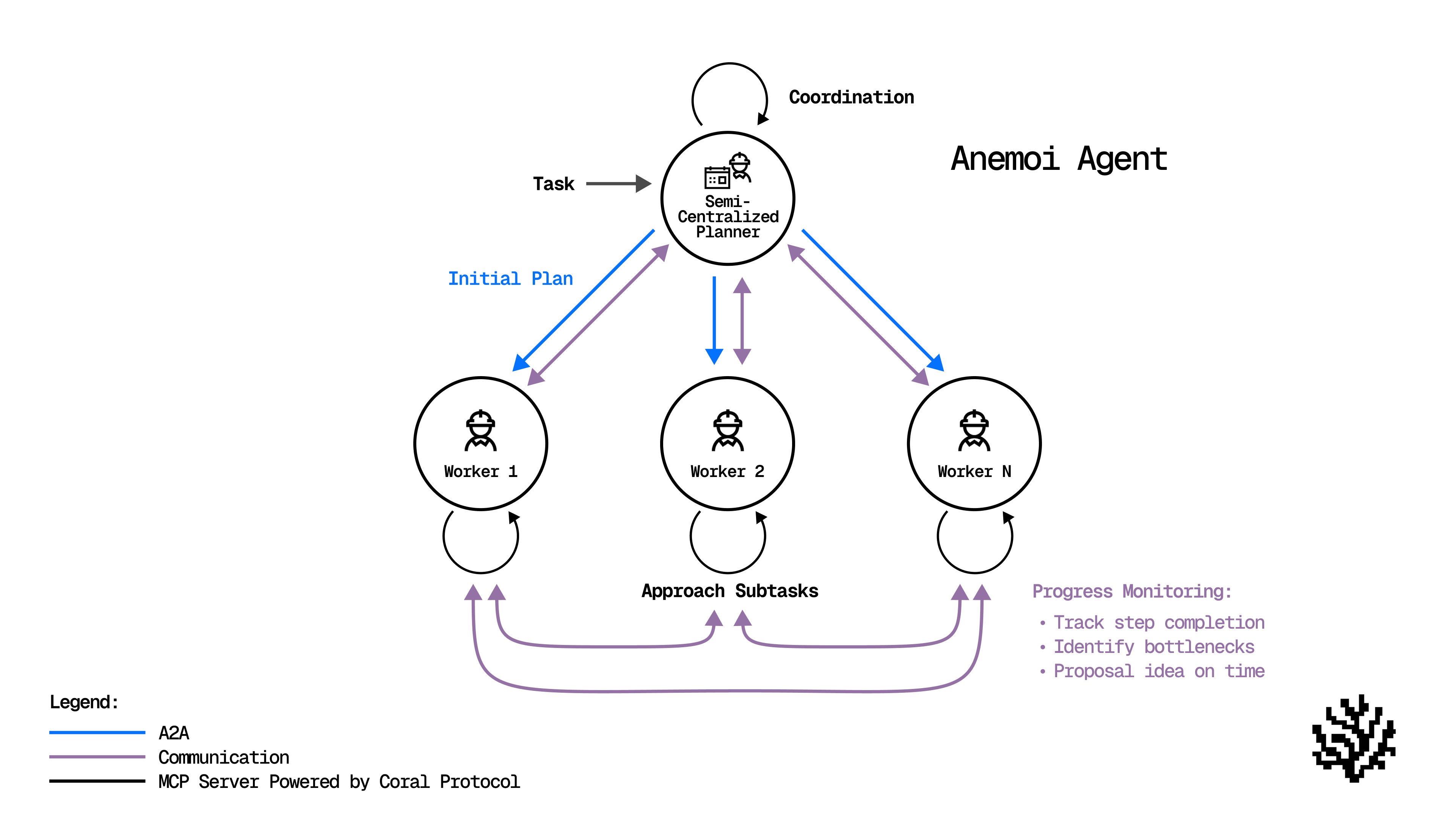}
    \caption{Architecture of the Anemoi: a semi-centralized multi-agent system based on the A2A communication MCP server from Coral Protocol.}
    \label{fig:architecture}
\end{figure}

\emph{Like winds connecting distant lands, \textbf{Anemoi} enables agents to communicate directly with one another in a semi-centralized network, achieving scalable coordination and seamless information flow.}

Large language models (LLMs) have demonstrated remarkable capabilities in a wide range of tasks, from text classification, natural language understanding~\cite{liu2025aiautomaticallyanalyzepublic}, and code generation~\cite{brown2020language, li2024autokaggle, zang2025kg}. However, when faced with complex, multi-step objectives that require diverse skills, a single LLM often struggles with maintaining context, managing long-horizon dependencies, and executing domain-specific actions efficiently~\cite{li2023camel, shi2025taskcraft}. To address these challenges, researchers have increasingly turned to MAS, where multiple specialized agents—each potentially powered by an LLM—collaborate to decompose, coordinate, and solve tasks~\cite{li2023camel, mialon2023gaia, xi2025rise}. This shift from single-model reasoning to distributed agent-based orchestration has opened new possibilities for scalability, modularity, and robustness in AI-driven problem-solving.

Recent advances in generalist MAS have been remarkable, with most existing designs following a context engineering plus centralized paradigm~\cite{fang2025cognitive, hu2025owl, tang2025agent}. In this setting, there is typically a centralized planner agent along with multiple worker agents responsible for specific tasks such as web search, file processing, or coding. Upon receiving a task, the planner agent decomposes it into subtasks and coordinates the worker agents to complete them. From a collaboration perspective, each agent essentially receives prompts and contextual information from a central or upstream planner, resulting in a predominantly unidirectional control flow. This paradigm presents two main limitations: 1) \textbf{Strong dependency on planner's capability:} If the planner is powered by a strong LLM, the system can perform well; however, replacing it with a small or mini LLM often leads to a significant drop in overall performance~\cite{mialon2023gaia}. 2) \textbf{Limited direct inter-agent communication:} In context-engineering-based designs, “collaboration” is typically achieved through prompt concatenation and manual context injection, without a dedicated channel for agents to directly exchange structured information\cite{yan2025beyond}. As a result, agents are unable to engage in genuine discussions that refine or adapt task plans over time. Collaboration is therefore reduced to one-shot context passing rather than iterative task improvement through interactive refinement.

Building on our observations of the current limitation of MAS, we propose the Anemoi, a semi-centralized generalist MAS underpinned by the \textbf{A2A communication model context protocol (MCP) ~\cite{anthropicMCP2024} server from Coral Protocol~\cite{georgio2025coralprotocolopeninfrastructure}}, as illustrated in Figure~\ref{fig:architecture}. The system combines a \textbf{semi-centralized planner agent} with multiple domain-specialized worker agents: the planner provide initial plan, while workers coordinate directly to monitor progress, resolve bottlenecks, and propose refinements, and all agents are allowed to communicate directly powered by the A2A communication MCP server. This paradigm yields three key benefits: \textbf{1)} it \textbf{reduces dependence on a centralized planner}, sustaining performance even with small LLMs; \textbf{2)} it supports \textbf{continuous plan updates} aligned with real-time execution; and \textbf{3)} it enables \textbf{genuine multi-agent debate for iterative task refinement}, leading to more reliable collaboration.

We evaluated the Anemoi on the GAIA benchmark~\cite{mialon2023gaia}, a challenging suite of real-world, multi-step tasks designed to assess the web-searching, multi-modal file processing, and coding capabilities of generalist AI systems. In our experiments, we adopted the same worker agent configurations as the current open-source State-of-the-Art (SOTA) system OWL~\cite{hu2025owl}, and used a small LLM (\texttt{GPT-4.1-mini}) as the planner agent alongside \texttt{GPT-4o} as the worker agents. Under this setting, our Anemoi achieved an accuracy of \textbf{52.73\%}, outperforming our reproduction of OWL with the same setup (43.63\%) by +9.09\%.

Our main contributions are summarized as follows:

\begin{enumerate}
    \item \textbf{A Semi-Centralized MAS Based on A2A Communication MCP server from Coral Protocol.} We propose the Anemoi, a semi-centralized multi-agent system built upon the A2A communication MCP server. This design eliminates the reliance on context engineering and the constraints of fully centralized coordination.
    
    \item \textbf{Strong Benchmark Performance with a Small Planner.} On the GAIA benchmark, Anemoi achieved 52.73\% accuracy even when the planner agent was powered by a small LLM (\texttt{GPT-4.1-mini}). Under identical LLM configurations, Anemoi outperformed the strongest open-source baseline OWL by +9.09\%.
    
\end{enumerate}

\section{Related Works}

\begin{table}[H]
\centering
\caption{Comparison of various multi-agent architectures. We summarize key differences across dimensions including pipelines of information flow, hierarchies of task plan, agent instantiation, and source availability.}
\label{tab:review}
\small
\begin{tabular}{@{} l l l l l @{}}
\toprule
\textbf{Agent} & \textbf{Flow Pipeline} & \textbf{Hierarchy} & \textbf{Instantiation} & \textbf{Availability} \\

\midrule
Agent KB & Context engineering & Centralized & Predefined & Open-source \\
Cognitive Kernel-Pro & Context engineering & Centralized & Predefined & Open-source \\
OWL & Context engineering & Centralized & Predefined & Open-source \\
\textbf{Our Anemoi} & \textbf{A2A communication} & \textbf{Semi-centralized} & Predefined & Open-source \\
\bottomrule

\end{tabular}
\end{table}

As agent technology has advanced, autonomous agent systems have evolved from relying on human-defined and fixed workflows~\cite{li2023camel, li2024autokaggle, ren2025simugen} to employing automated planning~\cite{fang2025cognitive, hu2025owl, qiu2025alita, tang2025agent}. This evolution marks a significant shift from single-task agents to general-purpose agents capable of solving complex, multi-stage tasks. Several representative architectures have emerged during this wave of development, as shown in Table~\ref{tab:review}.

Agent KB manages information flow through context engineering and utilizes a centralized shared memory pool with a knowledge base to guide its decision-making, laying the groundwork for the planning capabilities of subsequent agents~\cite{tang2025agent}. Building on this, Cognitive Kernel-Pro also adopts a centralized planning approach but introduces more sophisticated mechanisms, such as a reflection mechanism executed exclusively by the planner agent and a voting mechanism to enhance task reliability~\cite{fang2025cognitive}. These features significantly improve the agent's robustness when executing complex tasks. OWL follows a similar paradigm of centralized planner with Supervised Fine-Tuning (SFT) to improve the performance~\cite{hu2025owl}. 
Together, these works have advanced the maturity of context-engineering-based centralized planning agents, paving the way for more sophisticated general-purpose agent systems. However, such systems remain constrained by two major limitations. First, their overall intelligence and efficiency are capped by the capabilities of the centralized planner’s LLMs. Therefore, the centralized planners need to rely on cutting-edge closed-source LLMs to ensure adequate performance, such as GPT-5, Grok 4, and Claude Opus 4. The API call costs of these models rise sharply with the volume and complexity of tasks, making it difficult to operate the system at scale and at high frequency. Second, context engineering incurs significant token overhead and redundancy, since agent “collaboration” is realized through prompt concatenation and manual context injection rather than direct communication. This not only increases inference costs but also introduces information loss and restricts scalability. To overcome these bottlenecks, we propose a semi-centralized MAS based on the A2A communication MCP server from Coral Protocol to improve performance, robustness, and economy.

\section{Methodology}

\subsection{A2A Communication MCP Server}
At the core of the \textbf{Anemoi} is a dedicated infrastructure enabling multi-agent coordination via a structured, thread-based communication server from Coral Protocol. Each agent connects through a specialized \textbf{MCP server}, which provides tools for agent discovery, thread management, and message exchange. In this architecture, a \emph{thread} is a structured conversation channel in which multiple agents (and optionally humans or external services) can participate. The MCP server exposes key operations, including:

\begin{itemize}
    \item \texttt{list\_agents} -- discover all registered agents in the system.
    \item \texttt{create\_thread} -- initiate a conversation context with specified participants.
    \item \texttt{add\_participant} / \texttt{remove\_participant} -- dynamically adjust the set of agents involved in a thread.
    \item \texttt{send\_message} -- broadcast or directly address a message to agents within a thread.
    \item \texttt{wait\_for\_mentions} -- receive notifications when an agent is explicitly addressed.
    \item \texttt{close\_thread} -- terminate a conversation and optionally record its outcome.
\end{itemize}

This thread-based design ensures contextual compartmentalization---messages stay within their conversation, preserving clarity and preventing cross-talk---while supporting directed queries and task delegation between agents.

\subsection{Agent}

\begin{figure}[h]
    \centering
    \includegraphics[width=1.0\linewidth]{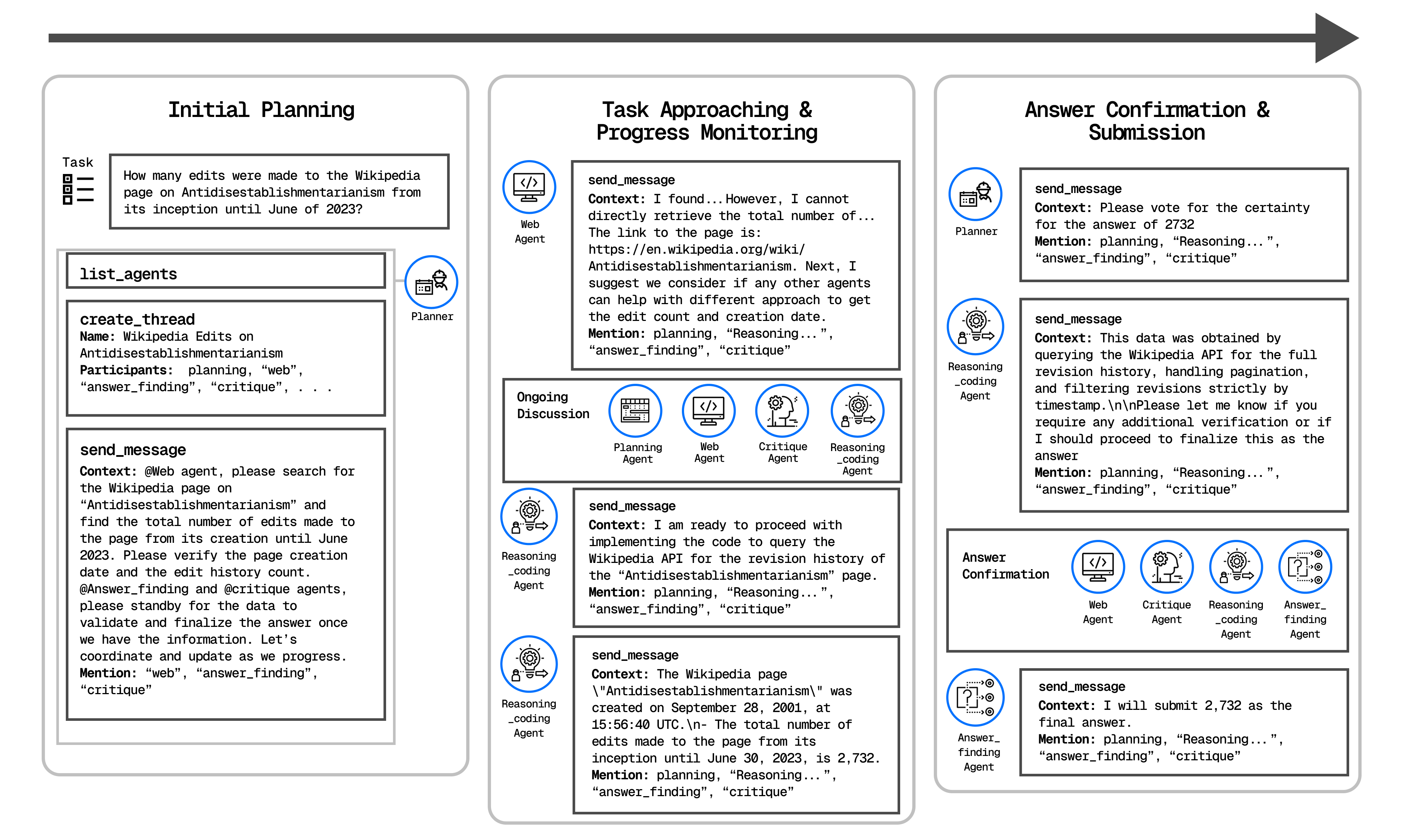}
    \caption{Overview of Anemoi. The system includes a planning agent to make initial plan, and a set of agents with different capability. The A2A communication MCP server enables all agents to monitor progress together.}
    \label{fig:workflow}
\end{figure}

As shown in Figure~\ref{fig:workflow}, the \textbf{Anemoi} framework consists of the following components:

\begin{itemize}
    \item \textbf{Planner Agent} -- Responsible for generating the initial plan at the beginning of a task and initiating coordination among the other agents to start execution.
    \item \textbf{Critique Agent} -- Continuously evaluates and questions the contributions of other agents to ensure the validity and certainty of their statements and conclusions.
    \item \textbf{Answer-Finding Agent} -- Compiles the final response based on validated outputs from other agents and submits the final answer.
    \item \textbf{Web Agent} -- Capable of performing web searches, extracting webpage content, simulating browser actions, and retrieving relevant online information.
    \item \textbf{Document Processing Agent} -- Processes a wide variety of local and remote documents, including PDF, DOCX, images, audio, and video files.
    \item \textbf{Reasoning \& Coding Agent} -- Specializes in reasoning, coding, and processing Excel files. It operates offline and cannot access the internet. If Python execution is required, it should be explicitly instructed to run the code after writing it.
\end{itemize}

Among these, the three worker agents (\textbf{Web Agent}, \textbf{Document Processing Agent}, and \textbf{Reasoning \& Coding Agent}) share the same configuration as those used in \textbf{OWL}, ensuring a fair experimental comparison. Notably, \textbf{each agent is integrated with the MCP toolkit provided by the A2A communication Server}, enabling them to monitor overall progress, track step completion, identify bottlenecks, and freely propose new ideas throughout task execution.

\subsection{Communication Pattern}

\begin{table}[H]
\centering
\caption{List of Symbols.}
\label{tab:symbol}
\small

\begin{tabular}{@{} ll @{}}

\toprule
\textbf{Symbol} & \textbf{Description} \\
\toprule
$\mathcal{A}$ & Set of all agents $\mathcal{A} = \{a_1, \ a_2, \ \dots \ , \ a_n\}$ \\
$\mathcal{W}$ & Set of worker agents, $\mathcal{W} = \{w_1, \ w_2, \ \dots \ , \ w_n\} \subset \mathcal{A}$ \\
$p$ & Planner agent \\
$c$ & Critique agent \\
$f$ & Answer-finding agent \\
$\tau$ & Communication thread created by MCP servers \\
$P_0$ & Set of initial participant agents, $P_0 = \{p_1, \ p_2, \ \dots \ , \ p_n\} \subseteq \mathcal{A}$ \\
$\pi_t$ & Task plan at step $t$ \\
$\phi$ & Task allocation mapping $\phi: \mathcal{W} \to \mathcal{T}$ \\
$r_w$ & Result produced by worker $w$ for $\phi(w)$ \\
$\texttt{critique}(r_w)$ & Evaluation outcome in $\{\texttt{accept}, \ \texttt{uncertain}\}$ \\
$o_i^{(t+1)}$ & Contribution of agent $a_i$ at time $(t+1)$ \\
$R^*$ & Final candidate solution proposed for consensus \\
$v_i(R^*)$ & Vote of agent $a_i$ on $R^*$ \\
$\mathcal{C}(\{v_i\})$ & Consensus aggregating votes \\
\bottomrule
\end{tabular}

\end{table}

Upon receiving a new task, the Anemoi follows a structured, semi-centralized communication workflow supported by the A2A MCP communication server (Table~\ref{tab:symbol} summarizes the key symbols used):

\begin{enumerate}[topsep=0pt]
    \item \textbf{Agent Discovery} \\
    Each agent $a_i \in \mathcal{A}$ invokes
    \[
        \{(name_j, desc_j)\}_{j=1}^n = \texttt{list\_agents}()
    \]
    where each tuple consists of the identifier and description of an available agent.

    \item \textbf{Thread Initialization} \\
    The planner agent $p$ creates a dedicated communication thread
    \[
        \tau = \texttt{create\_thread}(P_0),
    \]
    with initial participants $P_0 \subseteq \mathcal{A}$. The planner then broadcasts an initial plan $\pi_0$ to all participants and defines a tentative subtask allocation mapping
    \[
        \phi : \mathcal{W} \to \mathcal{T},
    \]
    where $\mathcal{W} \subset \mathcal{A}$ is the set of worker agents and $\mathcal{T}$ the set of subtasks.

    \item \textbf{Task Execution and Monitoring} \\
    Each worker $w \in \mathcal{W}$ executes its assigned subtask $\phi(w)$, producing a result
    \[
        r_w = w(\phi(w)).
    \]
    The critique agent evaluates each result through
    \[
        \texttt{critique}(r_w) \in \{\texttt{accept}, \texttt{uncertain}\}.
    \]
    Meanwhile, each agent $a_i \in P_0$ may generate a contribution to the ongoing discussion:
    \[
        o_i^{(t+1)} = a_i(\pi_t, \{r_w\}_{w \in \mathcal{W}}),
    \]
    where $o_i^{(t+1)}$ can represent a progress assessment, a critique, an alternative suggestion, or a revised plan proposal. These contributions collectively inform the evolving task context and may trigger updates to the plan.

    \item \textbf{Consensus Before Submission} \\
    Once a candidate solution $R^*$ emerges, the planner consults all participating agents for feedback:
    \[
        v_i(R^*) \in \{\texttt{approve}, \texttt{reject}\}, \quad \forall a_i \in P_0.
    \]
    The final decision is aggregated by a consensus conclusion $\mathcal{C}(\{v_i\})$.

    \item \textbf{Answer Submission} \\
    The answer-finding agent $f$ compiles and submits the validated result:
    \[
        \texttt{submit}(R^*).
    \]
\end{enumerate}

All message flows above are realized by the MCP primitives:
\[
    \texttt{send\_message}(\tau, m), \quad \texttt{wait\_for\_mentions}(\tau, a_i).
\]

This communication pattern offers several key advantages. 
\textbf{1)} It reduces dependency on a centralized planner: when progress deviates from expectations, multiple agents can collaboratively propose alternative solutions instead of relying solely on the planner’s reasoning. 
\textbf{2)} It supports adaptive plan refinement: in complex multi-step tasks, later subtasks often depend heavily on the accuracy of earlier results; as in trajectory prediction, initial steps tend to be more reliable, while later ones are more prone to drift. Unlike traditional centralized approaches that rigidly follow the initial plan, our method continuously updates the plan in response to real-time progress, ensuring that subsequent steps are re-aligned with the evolving task context. 
\textbf{3)} It enables multi-agent debate, allowing agents to engage in iterative discussions similar to human collaboration. Through this mechanism, agents can refine intermediate outputs, challenge assumptions, and incrementally converge toward more reliable solutions.

\section{Experiments}

\subsection{Baselines}
To evaluate the performance of the Anemoi, we compare it against a diverse set of both proprietary and open-source generalist multi-agent frameworks, covering a range of coordination paradigms and implementation strategies. The proprietary baselines include \textbf{DRP-val-v1.0}, \textbf{Omne}, and \textbf{Barcelona v0.1}, while the open-source baselines consist of \textbf{FRIDAY}~\cite{wu2024oscopilotgeneralistcomputeragents}, \textbf{Multi-Agent Exp v0.1}~\cite{microsoftGAIA2024}, \textbf{HuggingFace Agents}~\cite{smolagents2025}, \textbf{Magnetic-One}~\cite{fourney2024magenticonegeneralistmultiagentsolving}, and \textbf{OWL}~\cite{hu2025owl}. Proprietary performance results are taken directly from the official GAIA leaderboard, whereas open-source results are either obtained from the leaderboard or reproduced under consistent evaluation settings when necessary.

\subsection{Implementation Details}
Since the primary objective of this work is to evaluate the effectiveness of the A2A-based semi-centralized paradigm in coordinating complex tasks, we ensured a fair comparison by adopting exactly the same worker agents, including both toolkits and prompts, as those used in OWL. Specifically, \textbf{we encapsulated the three worker agents from OWL as tools}, and integrated them into three agents—each equipped with the A2A MCP server and powered by \texttt{GPT-4o}, to serve as our worker agents. Given our intuition that the advantages of the A2A-based semi-centralized paradigm would be more pronounced when the planner agent is powered by a weaker LLM, we employed \texttt{GPT-4.1-mini} as the LLM for the planner agent, while retaining \texttt{GPT-4o} for the worker agents. We did not include experiments with \texttt{GPT-4o-mini}, as preliminary tests indicated that the model lacked sufficient capability to reliably understand and correctly use the toolkits provided by the A2A MCP server. Since the original OWL paper does not report results under this setting (i.e., \texttt{GPT-4.1-mini} as the planner), we reproduced OWL accordingly. The original implementation of OWL was built using the CAMEL toolkit (version 0.2.46); however, some toolkits in this version are no longer available. We therefore used the corresponding toolkits from CAMEL version 0.2.70. Importantly, Anemoi and our reproduced OWL share exactly the same tools and model configurations, ensuring that any observed performance differences can be attributed solely to the coordination paradigm.

\begin{table}[H]
\centering
\setlength{\tabcolsep}{3pt}
\caption{Performance comparison of agent frameworks on GAIA validation set with accuracy score (\%) as the evaluation metric. Scores of open-source and proprietary frameworks were obtained from the official leaderboard. Specifically, OWL-rep is our reproduced result. The best-performing proprietary and open-source frameworks are highlighted in bold.}
\label{tab:gaia_performance}
\begin{tabular}{llcccc}
\toprule
\textbf{Agent Name} & \textbf{Base Model} & \textbf{Level 1} & \textbf{Level 2} & \textbf{Level 3} & \textbf{Average} \\
\midrule
\multicolumn{6}{c}{\textit{Proprietary Frameworks}} \\
\cmidrule(lr){1-6}
DRP-val-v.1.0 & - & 56.60 & 48.84 & 15.38 & 46.06 \\
omne & O1-Preview & 60.38 & 44.19 & 23.08 & 46.06 \\
Barcelona v0.1 & Claude-3.5-Sonnet & 62.26 & 50.00 & \textbf{26.92} & 50.30 \\
\midrule
\multicolumn{6}{c}{\textit{Open-Source Frameworks}} \\
\cmidrule(lr){1-6}
FRIDAY~\cite{wu2024oscopilotgeneralistcomputeragents} & GPT-4-Turbo & 45.28 & 34.88 & 11.54 & 34.55 \\
Multi-Agent Exp v0.1~\cite{microsoftGAIA2024} & GPT-4-Turbo & 54.72 & 38.37 & 11.54 & 39.39 \\
HuggingFace Agents~\cite{smolagents2025} & GPT-4o & 58.49 & 43.02 & 19.23 & 44.24 \\
Magnetic-One~\cite{fourney2024magenticonegeneralistmultiagentsolving} & O1 & 56.60 & 46.51 & 23.08 & 46.06 \\
OWL~\cite{hu2025owl} (pass@3) & GPT-4o-mini, GPT-4o & 64.15 & 45.34 & 19.23 & 47.27 \\
OWL-rep (pass@3) & GPT-4.1-mini, GPT-4o & 61.54 & 42.53 & 11.54 & 43.64 \\
\midrule
\multicolumn{6}{c}{\textit{Ours}} \\
\cmidrule(lr){1-6}
Anemoi (pass@3) & GPT-4.1-mini, GPT-4o & \textbf{71.70} & \textbf{52.33} & 15.38 & \textbf{52.73} \\
\bottomrule
\end{tabular}
\end{table}

\subsection{Main Results}

As shown in Table~\ref{tab:gaia_performance}, under the pass@3 setting, Anemoi achieves an accuracy of 52.73\% on the GAIA validation set, outperforming OWL with \texttt{GPT-4o-mini} and our reproduced OWL with \texttt{GPT-4.1-mini} by +5.46 and +9.09 percentage points, respectively. Notably, we also observe that even with a weaker planner model, Anemoi surpasses the performance of multiple proprietary and open-source frameworks that employ stronger LLMs. We attribute this improvement to our A2A communication + semi-centralized paradigm, which avoids the redundancy and token overhead inherent in context-engineering-based coordination, while enabling all agents to simultaneously track task progress. Regarding the lower accuracy of our reproduced \texttt{GPT-4.1-mini} OWL compared to the \texttt{GPT-4o-mini} version reported in the original paper, the performance gap can be explained by toolkit limitations. Specifically, the CAMEL v0.2.46 toolkit used in the original implementation included OpenAI’s \texttt{Whisper model}~\cite{openaiWhisper2022} for extracting transcripts from audio and video inputs. In CAMEL v0.2.70, \texttt{Whisper} support was removed, preventing accurate transcript extraction for audio tasks and making it impossible to process video audio tracks. We emphasize that our implementation of Anemoi also excludes \texttt{Whisper}, ensuring a fair comparison.

\section{Discussion}

\subsection{Comparative Analysis of Tasks Solved by Anemoi and OWL}

\begin{figure}[t]
    \centering
    \includegraphics[width=1\textwidth]{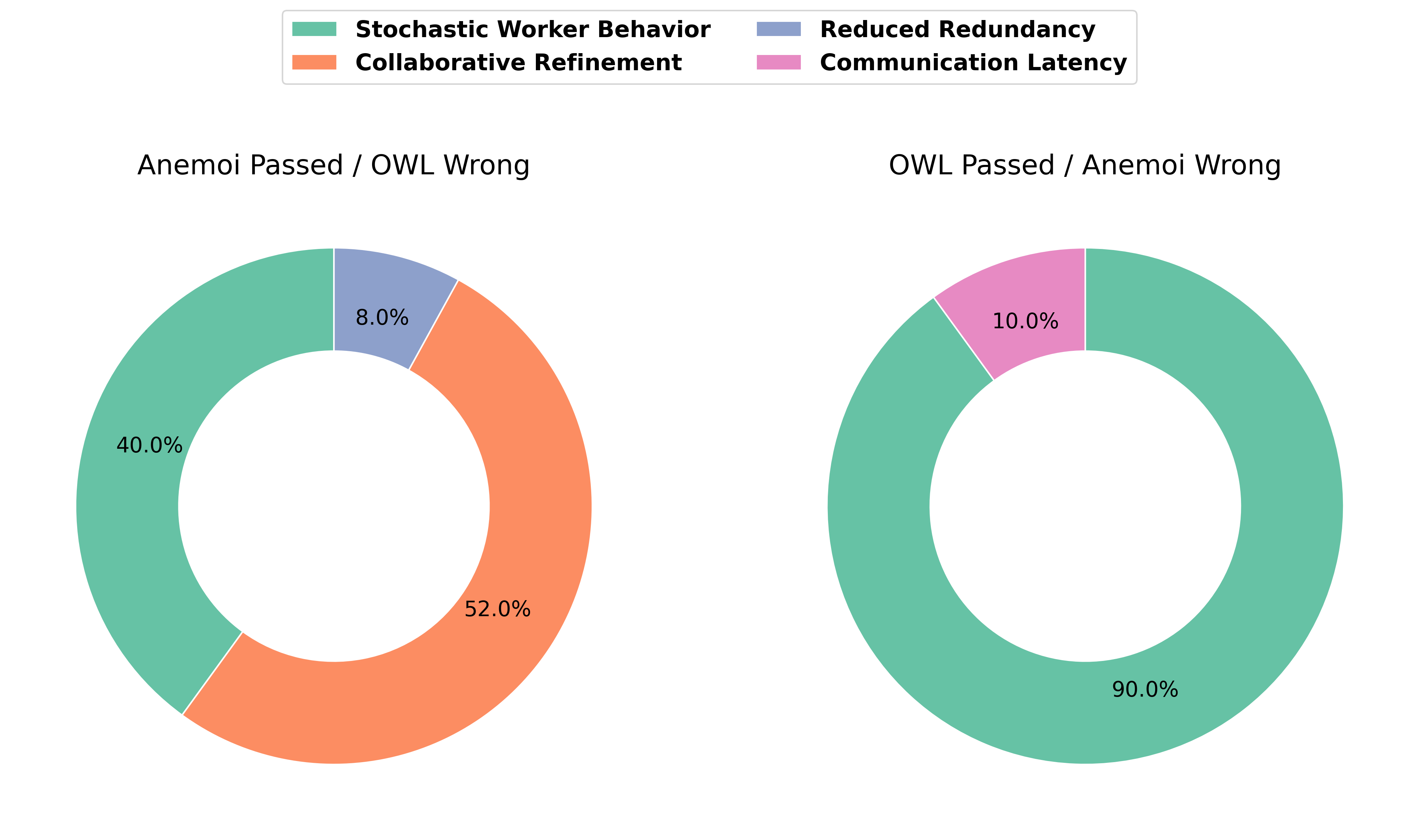}
    \caption{Comparison of task attribution categories between Anemoi and OWL. 
    The donut chart illustrates the distribution of reasons why Anemoi succeeded where OWL failed, and vice versa.}
    \label{fig:Anemoi_vs_owl_donut}
\end{figure}

Anemoi successfully solved 25 tasks that OWL failed to answer, while OWL solved 10 tasks that Anemoi did not. As illustrated in Figure~\ref{fig:Anemoi_vs_owl_donut}, the majority of Anemoi’s additional successes can be attributed to \textbf{collaborative refinement under the semi-centralized paradigm} (52\%). A smaller fraction arose from \textbf{reduced context redundancy enabled by the A2A communication protocol} (8\%), while the remaining cases were due to \textbf{stochastic worker behavior} (40\%), i.e., randomness in worker agents’ toolkit selection or usage.

Conversely, among the 10 tasks solved by OWL but not Anemoi, 90\% of the failures on Anemoi’s side were again due to \textbf{stochastic worker behavior}, while the remaining 10\% resulted from \textbf{communication latency in the web agent}, which failed to respond in time, because it was busy in executing task, forcing other agents to bypass it but ultimately preventing access to the correct information. \textit{Further details and case-level examples can be found in Appendix~\ref{sec:case_study_Anemoi_vs_owl}.}

\subsection{Error Analysis of Anemoi}

\begin{figure}[t]
    \centering
    \includegraphics[width=0.5\textwidth]{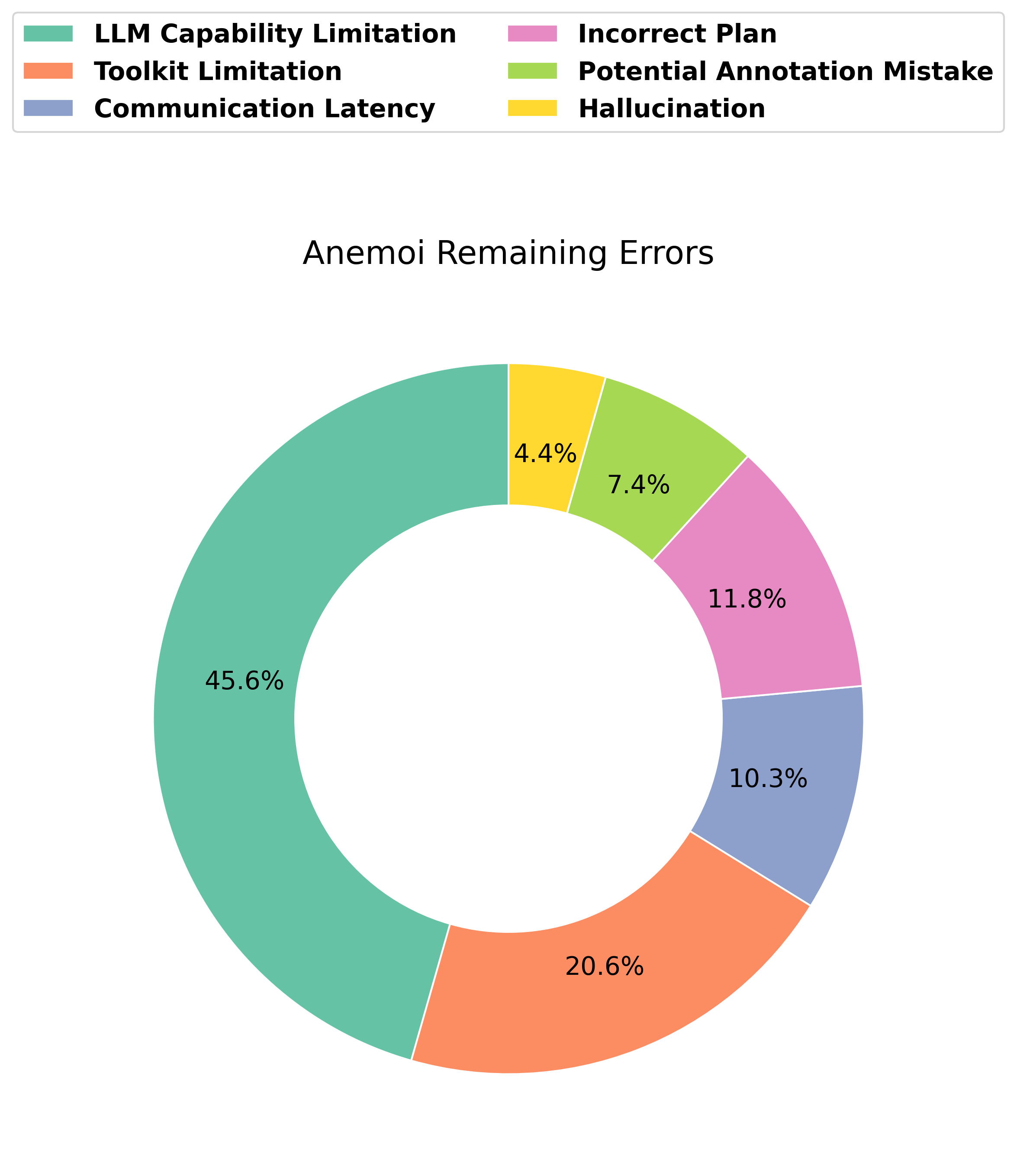}
    \caption{Remaining errors of the Anemoi. }
    \label{fig:Anemoi_remaining_errors}
\end{figure}

Beyond the comparative analysis with OWL, Anemoi still exhibits a total of 68 remaining errors. As shown in Figure~\ref{fig:Anemoi_remaining_errors}, the largest source of error is attributed to \textbf{LLM capability limitations} (45.6\%), where the LLM failed to select the correct toolkit or misused an available toolkit. Another 20.6\% of errors are due to \textbf{toolkit limitations}, i.e., inherent constraints or missing functionalities in the provided tools. 

A further 11.8\% of errors are caused by \textbf{incorrect plans}. Since Anemoi has the ability to update plans dynamically according to task progress, we only count the final updated plan if it remains incorrect. Errors caused by \textbf{communication latency} account for 10.3\%; these are primarily due to the web agent occasionally taking excessive time to complete search tasks, during which it cannot respond to other agents’ queries. This forces other agents to bypass the web agent and attempt alternative strategies, which ultimately fail. \textit{Additional case-level example of such failures are provided in Appendix~\ref{sec:failure_modes_Anemoi}.}

In addition, 7.4\% of the errors can be attributed to \textbf{potential annotation mistakes} in the benchmark, while the remaining 4.4\% result from \textbf{hallucinations} of the LLM.

\section{Conclusion}

In this work, we introduced the Anemoi, a novel semi-centralized MAS built upon the A2A communication MCP server from Coral Protocol. By reducing reliance on a single planner, supporting adaptive plan updates, and minimizing redundant context passing, our design enables more scalable execution. On the GAIA benchmark, Anemoi achieved {52.73\%} accuracy with a small LLM (\texttt{GPT-4.1-mini}) as the planner, surpassing OWL by +9.09\% under the same LLM configuration. This result not only highlights the effectiveness of semi-centralized A2A communication for generalist MAS, but also represents a concrete step toward realizing our broader vision of an \textit{Internet of Agents}.

\section{Acknowledgements}
We would like to express our sincere gratitude to Mustafa, Séafra Forder, and Mark Ruben I. Abacajan for their valuable support and insightful suggestions throughout the development of this work. We also gratefully acknowledge the funding support provided by Coral Protocol for this project.


\bibliographystyle{plain}
\bibliography{ref}

\begin{thebibliography}{10}

\bibitem{anthropicMCP2024}
Anthropic.
\newblock Introducing the model context protocol, November 25 2024.
\newblock Accessed: 2025-08-18.

\bibitem{brown2020language}
Tom Brown, Benjamin Mann, Nick Ryder, Melanie Subbiah, Jared~D Kaplan, Prafulla
  Dhariwal, Arvind Neelakantan, Pranav Shyam, Girish Sastry, Amanda Askell,
  et~al.
\newblock Language models are few-shot learners.
\newblock {\em Advances in neural information processing systems},
  33:1877--1901, 2020.

\bibitem{fang2025cognitive}
Tianqing Fang, Zhisong Zhang, Xiaoyang Wang, Rui Wang, Can Qin, Yuxuan Wan,
  Jun-Yu Ma, Ce~Zhang, Jiaqi Chen, Xiyun Li, et~al.
\newblock Cognitive kernel-pro: A framework for deep research agents and agent
  foundation models training.
\newblock {\em arXiv preprint arXiv:2508.00414}, 2025.

\bibitem{fourney2024magenticonegeneralistmultiagentsolving}
Adam Fourney, Gagan Bansal, Hussein Mozannar, Cheng Tan, Eduardo Salinas,
  Erkang Zhu, Friederike Niedtner, Grace Proebsting, Griffin Bassman, Jack
  Gerrits, Jacob Alber, Peter Chang, Ricky Loynd, Robert West, Victor Dibia,
  Ahmed Awadallah, Ece Kamar, Rafah Hosn, and Saleema Amershi.
\newblock Magentic-one: A generalist multi-agent system for solving complex
  tasks.
\newblock {\em arXiv preprint arXiv:2411.04468}, 2024.

\bibitem{georgio2025coralprotocolopeninfrastructure}
Roman~J. Georgio, Caelum Forder, Suman Deb, Andri Rahimov, Peter Carroll, and
  {\"O}nder G{\"u}rcan.
\newblock Coral protocol: Open infrastructure connecting the internet of
  agents.
\newblock {\em arXiv preprint arXiv:2505.00749}, 2025.

\bibitem{hu2025owl}
Mengkang Hu, Yuhang Zhou, Wendong Fan, Yuzhou Nie, Bowei Xia, Tao Sun, Ziyu Ye,
  Zhaoxuan Jin, Yingru Li, Qiguang Chen, et~al.
\newblock Owl: Optimized workforce learning for general multi-agent assistance
  in real-world task automation.
\newblock {\em arXiv preprint arXiv:2505.23885}, 2025.

\bibitem{li2023camel}
Guohao Li, Hasan Hammoud, Hani Itani, Dmitrii Khizbullin, and Bernard Ghanem.
\newblock Camel: Communicative agents for" mind" exploration of large language
  model society.
\newblock {\em Advances in Neural Information Processing Systems},
  36:51991--52008, 2023.

\bibitem{li2024autokaggle}
Ziming Li, Qianbo Zang, David Ma, Jiawei Guo, Tuney Zheng, Minghao Liu, Xinyao
  Niu, Yue Wang, Jian Yang, Jiaheng Liu, et~al.
\newblock Autokaggle: A multi-agent framework for autonomous data science
  competitions.
\newblock {\em arXiv preprint arXiv:2410.20424}, 2024.

\bibitem{liu2025aiautomaticallyanalyzepublic}
Jing Liu, Xinxing Ren, Yanmeng Xu, and Zekun Guo.
\newblock Can ai automatically analyze public opinion? a llm agents-based
  agentic pipeline for timely public opinion analysis.
\newblock {\em arXiv preprint arXiv:2505.11401}, 2025.

\bibitem{mialon2023gaia}
Gr{\'e}goire Mialon, Cl{\'e}mentine Fourrier, Thomas Wolf, Yann LeCun, and
  Thomas Scialom.
\newblock Gaia: a benchmark for general ai assistants.
\newblock In {\em The Twelfth International Conference on Learning
  Representations}, 2023.

\bibitem{microsoftGAIA2024}
Microsoft.
\newblock Multi-agent experiment v0.1 msr ai frontiers (autogen team members),
  2024.
\newblock Accessed: 2025-08-18.

\bibitem{openaiWhisper2022}
OpenAI.
\newblock Whisper: Robust speech recognition via large-scale weak supervision
  (github repository), 2022.

\bibitem{qiu2025alita}
Jiahao Qiu, Xuan Qi, Tongcheng Zhang, Xinzhe Juan, Jiacheng Guo, Yifu Lu, Yimin
  Wang, Zixin Yao, Qihan Ren, Xun Jiang, et~al.
\newblock Alita: Generalist agent enabling scalable agentic reasoning with
  minimal predefinition and maximal self-evolution.
\newblock {\em arXiv preprint arXiv:2505.20286}, 2025.

\bibitem{ren2025simugen}
Xinxing Ren, Qianbo Zang, and Zekun Guo.
\newblock Simugen: Multi-modal agentic framework for constructing block
  diagram-based simulation models.
\newblock {\em arXiv preprint arXiv:2506.15695}, 2025.

\bibitem{smolagents2025}
Aymeric Roucher, Albert~Villanova del Moral, Thomas Wolf, Leandro von Werra,
  and Erik Kaunism{\"a}ki.
\newblock {\textquotesingle}smolagents{\textquotesingle}: A smol library to
  build great agentic systems, 2025.
\newblock Accessed: 2025-08-18.

\bibitem{shi2025taskcraft}
Dingfeng Shi, Jingyi Cao, Qianben Chen, Weichen Sun, Weizhen Li, Hongxuan Lu,
  Fangchen Dong, Tianrui Qin, King Zhu, Minghao Liu, et~al.
\newblock Taskcraft: Automated generation of agentic tasks.
\newblock {\em arXiv preprint arXiv:2506.10055}, 2025.

\bibitem{tang2025agent}
Xiangru Tang, Tianrui Qin, Tianhao Peng, Ziyang Zhou, Daniel Shao, Tingting Du,
  Xinming Wei, Peng Xia, Fang Wu, He~Zhu, et~al.
\newblock Agent kb: Leveraging cross-domain experience for agentic problem
  solving.
\newblock {\em arXiv preprint arXiv:2507.06229}, 2025.

\bibitem{wu2024oscopilotgeneralistcomputeragents}
Zhiyong Wu, Chengcheng Han, Zichen Ding, Zhenmin Weng, Zhoumianze Liu, Shunyu
  Yao, Tao Yu, and Lingpeng Kong.
\newblock Os-copilot: Towards generalist computer agents with self-improvement.
\newblock {\em arXiv preprint arXiv:2402.07456}, 2024.

\bibitem{xi2025rise}
Zhiheng Xi, Wenxiang Chen, Xin Guo, Wei He, Yiwen Ding, Boyang Hong, Ming
  Zhang, Junzhe Wang, Senjie Jin, Enyu Zhou, et~al.
\newblock The rise and potential of large language model based agents: A
  survey.
\newblock {\em Science China Information Sciences}, 68(2):121101, 2025.

\bibitem{yan2025beyond}
Bingyu Yan, Zhibo Zhou, Litian Zhang, Lian Zhang, Ziyi Zhou, Dezhuang Miao,
  Zhoujun Li, Chaozhuo Li, and Xiaoming Zhang.
\newblock Beyond self-talk: A communication-centric survey of llm-based
  multi-agent systems.
\newblock {\em arXiv preprint arXiv:2502.14321}, 2025.

\bibitem{zang2025kg}
Qianbo Zang, Christophe Zgrzendek, Igor Tchappi, Afshin Khadangi, and Johannes
  Sedlmeir.
\newblock Kg-htc: Integrating knowledge graphs into llms for effective
  zero-shot hierarchical text classification.
\newblock {\em arXiv preprint arXiv:2505.05583}, 2025.

\end{thebibliography}


\appendix

\section{Case Study: Anemoi Success vs. OWL Failure}
\label{sec:case_study_Anemoi_vs_owl}

\begin{purposebox}{Task}
\textbf{Task:} \textit{How many edits were made to the Wikipedia page on \texttt{Antidisestablishmentarianism} from its inception until June of 2023?} \\

\textbf{Ground Truth Answer:} 2732
\end{purposebox}

\begin{purposebox}{Anemoi's Final Answer}
The Anemoi produced the following final response:

\textit{"2732"}
\end{purposebox}

\begin{purposebox}{Anemoi's Agent Dialogue}
\small\ttfamily  
\textbf{[Planner]}: \\
Web agent, please search for the Wikipedia page on "Antidisestablishmentarianism" and find the total number of edits made to the page from its creation until June 2023. Please verify the page creation date and the edit history count. Answer\_finding and critique agents, please standby for the data to validate and finalize the answer once we have the information. Let's coordinate and update as we progress. \\[6pt]

\textbf{[Web\_Agent]}: \\
Starting search for the Wikipedia page on "Antidisestablishmentarianism" to find the total number of edits made from its inception until June 2023. I will verify the page creation date and the edit history count from the page's history or relevant metadata. \\[6pt]

\textbf{[Web\_Agent]}: \\
I found the Wikipedia page for "Antidisestablishmentarianism" and located the link to its edit history. However, I cannot directly retrieve the total number of edits or the page creation date programmatically. The user or another agent can manually check the "View history" tab on the Wikipedia page to find the total number of edits and the creation date. The link to the page is: \url{https://en.wikipedia.org/wiki/Antidisestablishmentarianism}. Next, I suggest we consider if any other agents can help with this manual extraction or if we should try a different approach to get the edit count and creation date. \\[6pt]

\textbf{[Planner (on going discussion)]}: \\
Thanks for the update, web agent. Since the web agent cannot directly retrieve the total number of edits or the page creation date programmatically, we need to consider alternative approaches: \\
1. The answer\_finding agent or critique agent could attempt to manually extract the information from the Wikipedia page's "View history" tab if they have browsing capabilities. \\
2. We could try to find a Wikipedia API or a third-party tool that provides the edit count and creation date for Wikipedia pages. \\
3. If none of the above is feasible, we may need to instruct the user to manually check the edit history and provide the data. \\
\textbf{Answer\_finding and critique agents, please advise on the best approach or if you can assist with manual extraction or API usage. Web agent, please standby for further instructions.} \\[6pt]

\textbf{[Critique\_Agent (on going discussion)]}: \\
Planning agent, thank you for the summary. Answer\_finding agent, given the options and the lack of direct programmatic access from the web agent, do you have browsing or API access capabilities to assist with manual extraction or querying Wikipedia APIs or third-party tools? If not, do you have suggestions on the most reliable third-party tools or APIs we could use to get the edit count and creation date with 100\% certainty? Web agent, please standby for further instructions. \\[6pt]

\textbf{[Answer\_finding\_Agent (on going discussion)]}: \\
To planning and critique agents: I do not have direct browsing capabilities \textbf{but can request the reasoning\_coding agent to write and execute code to query the Wikipedia API for the revision history of the "Antidisestablishmentarianism" page.} This code can: retrieve all revisions with timestamps; filter revisions up to June 30, 2023; count the total number of edits; identify the earliest revision timestamp as the page creation date. This approach should provide a precise and verifiable count of edits up to the specified date. Critique and planning agents, do you agree with this approach? If so, I will proceed to coordinate with the reasoning\_coding agent to implement this solution. \\[6pt]

\textbf{[Planner (on going discussion)]}: \\
Answer finding agent, I confirm that adding the reasoning\_coding agent to this thread to assist with the implementation and coding of the Wikipedia API query approach is a good next step. This will help ensure the code is written with all necessary considerations for pagination, filtering, and anomaly handling to achieve 100\% certainty. \\[6pt]

\textbf{[Reasoning\_Coding\_Agent (on going discussion)]}: \\
Hello team, I am ready to proceed with implementing the code to query the Wikipedia API for the revision history of the "Antidisestablishmentarianism" page. I will ensure to handle pagination, filter revisions strictly by timestamp up to June 30, 2023, count the total edits, and identify the earliest revision timestamp as the page creation date. I will also consider potential anomalies such as page moves or merges by verifying the earliest revision timestamp and ensuring the data corresponds to the current page title. I will start writing the code now and will update you with the results shortly. \\[6pt]

\textbf{[Reasoning\_Coding\_Agent]}: \\
Hello answer\_finding and critique agents, I have completed the code execution to query the Wikipedia API for the revision history of the "Antidisestablishmentarianism" page. Results: \\
- Page creation date: September 28, 2001, 15:56:40 UTC \\
- Total number of edits up to June 30, 2023: 2,732 \\
This count includes all revisions up to and including June 30, 2023, and the earliest revision timestamp was used to determine the page creation date. Please review and advise on the next steps for validation and finalization of the answer. \\
\end{purposebox}

\begin{purposebox}{OWL's Final Answer}
The OWL framework produced the following final response:

\textit{"Unable to determine due to incomplete data"}
\end{purposebox}

\begin{purposebox}{OWL's Agent Dialogue}
\small\ttfamily  
\textbf{[Planner]}: \\
\quad <tasks> \\
\quad \ \ <task>Search for the Wikipedia page on "Antidisestablishmentarianism" and identify the page's inception date (the date when the page was first created).</task> \\
\quad \ \ <task>Retrieve the complete revision history of the Wikipedia page on "Antidisestablishmentarianism" from its inception until June 2023, including the total number of edits made.</task> \\
\quad \ \ <task>Verify the total number of edits by cross-checking with an alternative source or method, such as archived versions or Wikipedia's revision statistics tools.</task> \\
\quad \ \ <task>Summarize the findings and provide the final answer to the question: "How many edits were made to the Wikipedia page on Antidisestablishmentarianism from its inception until June of 2023?"</task> \\
\quad </tasks> \\[6pt]

\textbf{[Coordinator\_Agent]}: \\
<task>Search for the Wikipedia page on "Antidisestablishmentarianism" and identify the page's inception date (the date when the page was first created).</task> \\

Assignee: Web\_Agent\\[6pt]

\textbf{[Web\_Agent]}: \\
The Wikipedia page for "Antidisestablishmentarianism" was first created on 18 May 2014. \\[6pt]

\textbf{[Coordinator\_Agent]}: \\
<task>Retrieve the complete revision history of the Wikipedia page on "Antidisestablishmentarianism" from its inception until June 2023, including the total number of edits made.</task> \\

Assignee: Web\_Agent\\[6pt]

\textbf{[Web\_Agent]}: \\
I attempted to retrieve the revision history of the Wikipedia page on "Antidisestablishmentarianism" up to June 2023, but it seems there is no direct access to the complete revision history through the current tool. \\

To proceed, I suggest using the Wikipedia API or a similar tool to programmatically access the revision history. This would allow us to count the total number of edits made to the page from its inception until June 2023. Alternatively, manually accessing the revision history through the Wikipedia page itself might be necessary if API access is not feasible. \\[6pt]

\textbf{[Planner (replan)]}: \\
\quad <tasks> \\
\quad \ \ <task>Identify the exact Wikipedia page title for "Antidisestablishmentarianism" and confirm the page's creation date (inception date) from Wikipedia.</task> \\
\quad \ \ <task>Retrieve the total number of edits made to the Wikipedia page on "Antidisestablishmentarianism" from its inception date until June 2023 by accessing the page's revision history or statistics using Wikipedia tools or API.</task> \\
\quad \ \ <task>Cross-verify the total number of edits obtained by checking alternative sources such as Wikipedia's revision statistics tools, archived versions, or other reliable revision tracking services.</task> \\
\quad \ \ <task>Summarize and provide the final answer to the question: "How many edits were made to the Wikipedia page on Antidisestablishmentarianism from its inception until June of 2023?" based on the verified data.</task> \\
\quad </tasks> \\[6pt]

\textbf{[Coordinator\_Agent]}: \\
<task>Identify the exact Wikipedia page title for "Antidisestablishmentarianism" and confirm the page's creation date (inception date) from Wikipedia.</task> \\

Assignee: Web\_Agent\\[6pt]

\textbf{[Web\_Agent]}: \\
I attempted to find the creation date of the Wikipedia page for "Antidisestablishmentarianism" by navigating through its revision history. However, due to the large number of revisions and the limitations on the number of rounds for browsing, I was unable to reach the earliest revision to confirm the exact creation date. \\

To proceed, I recommend using a more efficient method to access the earliest revision, such as directly querying Wikipedia's API for the first revision of the page or using a tool that can handle large datasets more effectively. This would allow us to determine the inception date of the page accurately. \\
\end{purposebox}

\section{Typical Failure Mode of Anemoi Agent}
\label{sec:failure_modes_Anemoi}

\begin{promptbox}{Task}
\textbf{Task:} \textit{How many nonindigenous crocodiles were found in Florida from the year 2000 through 2020? You can get the data from the USGS Nonindigenous Aquatic Species database.} \\

\textbf{Ground Truth Answer:} 6
\end{promptbox}

\begin{promptbox}{Anemoi's Final Answer}
The Anemoi produced the following final response:

\textit{"give up: The web agent has not provided any data or updates from the USGS Nonindigenous Aquatic Species database regarding nonindigenous crocodiles found in Florida from 2000 through 2020 despite multiple requests and reminders. Without this data, we cannot determine the count with any certainty. Therefore, we must give up on this task due to lack of data availability and inability to verify the answer."}
\end{promptbox}

\begin{promptbox}{Anemoi's Agent Dialogue}
\small\ttfamily  
\textbf{[Planner]}: \\
I suggest the following initial plan: 1. Use the web agent to access the USGS Nonindigenous Aquatic Species database and locate data on crocodiles in Florida. 2. Extract the relevant data for the years 2000 through 2020. 3. Aggregate the count of nonindigenous crocodiles found in that timeframe. 4. Verify the data and format the final answer as a simple number. Please provide your input or suggestions before we proceed. \\[6pt]

\textbf{[Web\_Agent]}: \\
Hello team, I will start by searching the USGS Nonindigenous Aquatic Species database for data on nonindigenous crocodiles found in Florida from 2000 through 2020. I will look for any relevant reports, datasets, or records that specify the count of such crocodiles in that timeframe and location. \\[6pt]

\textbf{[Planner]}: \\
Web agent, could you please provide an update on your progress in locating the data on nonindigenous crocodiles found in Florida from 2000 through 2020 in the USGS Nonindigenous Aquatic Species database? If you have found relevant data or sources, please share them with the team so we can proceed with extraction and analysis. If you need any assistance or clarification, please let us know. \\[6pt]

\textbf{[Answer\_finding\_Agent]}: \\
@web, we have not yet received the data or findings from the USGS Nonindigenous Aquatic Species database regarding nonindigenous crocodiles in Florida from 2000 to 2020. Could you please provide an update on your progress or any difficulties encountered? If you need specific guidance on how to extract or summarize the data, please let us know so we can assist. \\[6pt]

\textbf{[Critique\_Agent]}: \\
Since there are no new updates yet from the web agent regarding the data retrieval, I want to reiterate the importance of documenting the data source URL or document, the exact counts, and any assumptions or data handling methods used once the data is found. Also, please provide any metadata or documentation that can help us assess the completeness and accuracy of the data. This will be critical for us to verify the final count and ensure 100\% certainty in our answer. If the web agent anticipates any delays or difficulties in finding the data, please communicate that so we can adjust our approach accordingly. \\[6pt]

... (After several rounds of consultation and discussions on alternative approaches) \\[6pt]

\textbf{[Critique\_Agent]}: \\
Hello @web, critique agent here again. We have been waiting for some time without updates on the data retrieval from the USGS Nonindigenous Aquatic Species database for nonindigenous crocodiles found in Florida from 2000 to 2020.  Could you please provide a status update or let us know if you are facing any technical difficulties or access issues? If so, please communicate immediately so we can assist or consider alternative approaches.\ Our goal remains to ensure 100\% certainty in the final count, so clear data source documentation and verification steps will be essential once data is available. Thank you for your cooperation. \\[6pt]
\end{promptbox}


\end{document}